\documentclass[prb,aps,twocolumn,showpacs,amsmath,amssymb,floatfix,citeautoscript]{revtex4}
\usepackage{graphicx}
\usepackage{dcolumn}
\usepackage{bm}
\usepackage{amsmath}
\usepackage{amssymb}
\usepackage[normal]{subfigure}

\makeatletter

\begin{document}

\title{Sensing Kondo correlations in a suspended carbon nanotube mechanical resonator with spin-orbit coupling}

\author{Dong E. Liu}
\affiliation{State Key Laboratory of Low-Dimensional Quantum Physics and Department of Physics, Tsinghua University, Beijing 100084, China}

\date{\today}

\begin{abstract}
We study electron mechanical coupling in a suspended carbon nanotube (CNT) quantum dot device.
Electron spin couples to the flexural vibration mode due to spin-orbit coupling in the
electron tunneling processes. In the weak coupling limit, i.e. electron-vibration
coupling is much smaller than electron energy scale, 
the damping and resonant frequency shift of the CNT resonator
can be obtained by calculating the dynamical spin susceptibility.
We find that strong spin-flip scattering processes in Kondo regime 
significantly affect the mechanical motion of the carbon nanotube: 
Kondo effect induces strong damping and frequency shift of the CNT resonator. 
\end{abstract}


\maketitle

\section{Introduction and short summary}

Carbon nanotubes (CNTs) have been considered as an ideal platform
for quantum dot (QD) devices \cite{Tans97,KouwenhovenReview}, which show
Coulomb blockade oscillations, Luttinger liquid behavior~\cite{CNT-Luttinger}, Kondo effects \cite{Nygard00,Jarillo-Herrero05},
and phase transitions \cite{Mebrahtu12,Wang552}.
CNTs is also emerging as a promising material for high-quality quantum nanomechanical 
applications \cite{Sapmaz03,Sazonova04,Garcia-Sanchez07,Huttel09,Lassagne09,Steele09,Moser13,MoserNN2014}
due to their low mass and high stiffnesses, and thus is useful in quantum sensing~\cite{AresPRL16,KhoslaPRX18,QuantumSensingRMP},
and  in quantum information processing~\cite{Palyi12,Ohm12,RipsPRL13,NoriPRL16}.
The strong coupling between mechanical vibrations and the electronic degree
of freedom was achieved in high quality suspended CNT QD resonators \cite{Lassagne09,Steele09},
when single electron tunnelings through the CNT QD are turned on. 
This electron-vibration coupling provides an opportunity to study the
electron correlations and quantum noise through the measurement of the vibration of CNT resonator,
and provides a way to achieve the electron-induced cooling of the resonator.
Indeed, strong damping, frequency shift, and their nonlinearity effects of the CNT resonator are observed \cite{Lassagne09,Steele09}. 

In those observations, the electron-vibration coupling 
is induced by gate capacitance dependence of the resonator displacement, which only 
results in the coupling between the electron density and the vibrations. 
Therefore, Kondo effects, caused by spin-flip scatterings due to strong effective
spin exchange coupling in low temperature \cite{HewsonBook}, seem to be irrelevant when considering mechanical effects.
However, a recent theoretical proposal \cite{Rudner&Rashba} shows that
the coupling between the flexural vibration mode and the electron spin can be 
achieved in CNT, because the spin-orbit coupling in CNT \cite{Ando00,Huertas-Hernando06,Kuemmeth08,SteeleNC13} 
tends to align the spin with the tangent direction of CNT.
Most interestingly, Kondo effects become relevant due to such spin-vibration coupling ($\Delta_{SO}$):
Strong spin-flip scattering processes in Kondo regime might significantly affect
the CNT vibrations. Therefore, CNT resonator may also provide a way to "sensing" those 
quantum many body correlations.

In this work, we study a suspended CNT QD coupled to source-drain leads in the Kondo
regime. Both the electron density-vibration coupling due to gate capacitance change
and the spin-vibration coupling due to spin-orbit coupling are considered.
When those couplings are much smaller than an electron energy scale (Kondo temperature),
a perturbation treatment shows that the damping $\gamma$ and resonant frequency
($\omega_0$) shift of the CNT resonator due to electron (both density and spin) vibration coupling
can be directly connected to the dynamical charge and spin susceptibilities.
In the experimental realizable regime $\omega_0\sim T_K^{SU(2)}\ll \Delta_{KK'}\ll\Delta_{SO}$ 
($\Delta_{KK'}$ is intervalley scattering), those dynamical susceptibilities can 
be obtained from non-crossing approximation \cite{Coleman84,BickersPRB87,Brunner&Langreth97,HewsonBook}.
We show that the strong spin-flip scatterings in Kondo regime induce strong damping and frequency 
shift of the CNT resonator. Those effects can be detected by using a finite frequency
noise measurement \cite{Moser13}.

\begin{figure}[t]
\centering
\includegraphics[width=3.2in,clip]{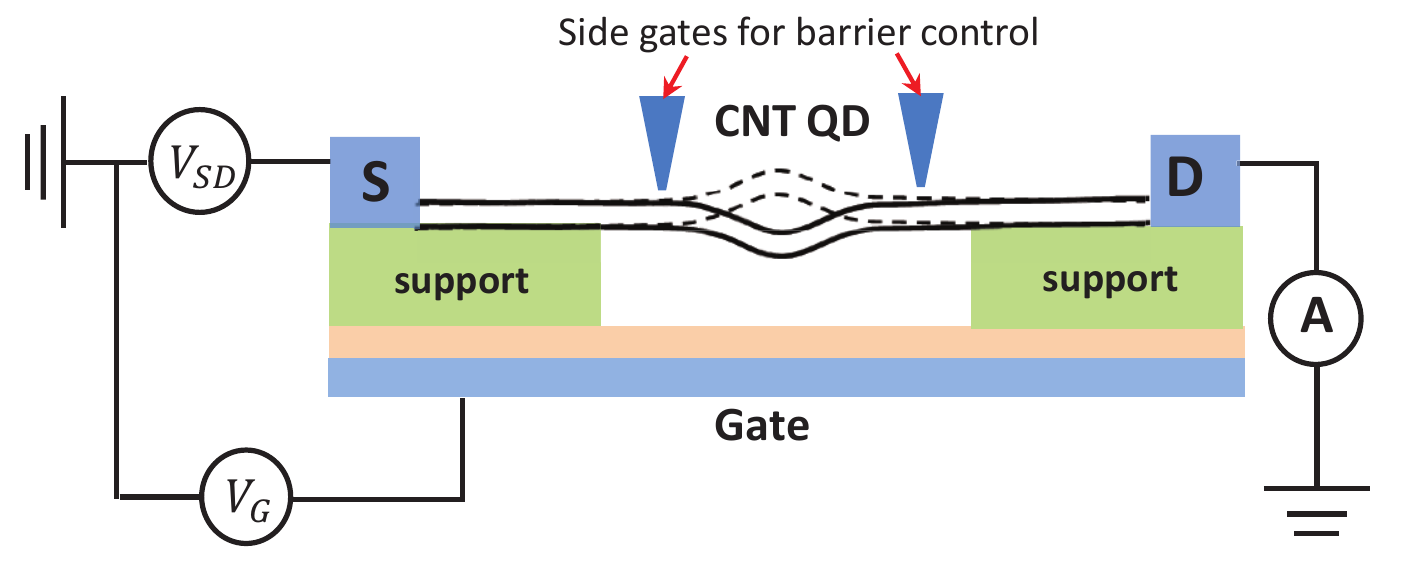}
\caption{(color online) Schematics of the experimetal setup:
a suspended doubly-clamped semiconducting CNT QD is connected by source and drain leads.
Back gate is used to control the energy level of the QD, and the two side gates are
for tunneling barrier control.
}
\label{fig:setup_KondoV}
\end{figure}

\section{Model and Hamiltonian}

We consider a suspended doubly-clamped semiconducting CNT QD
shown in Fig. \ref{fig:setup_KondoV}.
Due to the twofold real spin and twofold isospin symmetries, all the eigenenergies of the CNT quantum
dot becomes fourfold degenerate \cite{Liang02,Buitelaar02}. For each eigen-energy, the eigenstates
can be written as $|\tau\, s\rangle=|\tau\rangle\otimes|s\rangle$,
where $|\tau\rangle=|\frac{1}{2}\rangle,\,|-\frac{1}{2}\rangle$ represents
the isospin and $|s\rangle=|\frac{1}{2}\rangle,\,|-\frac{1}{2}\rangle$
represents the real spin. The CNT also includes a spin-orbit
coupling term and an intervalley scattering term, and
the Hamiltonian is as follows
\begin{equation}
H_{{\rm CNT}}=H_{0}+\frac{\Delta_{SO}}{2}\tau_{3}(\textbf{s}\cdot\hat{\textbf{t}})+\Delta_{KK'}\tau_{1}
\end{equation}
where $H_{0}|\tau\, s\rangle=E_{0}|\tau\, s\rangle$,  $\Delta_{SO}$ and $\Delta_{KK'}$ are 
the spin-orbit coupling and the intervalley scattering, respectively. 
$\hat{\textbf{t}}=(t_{x},t_{y},t_{z})$ is the local tangent vector along the CNT axis,
and we assume the CNT is along the $z$-direction without deformation. For the vibration, we focus
on the lowest flexural mode, which can be described by a harmonic oscillator
\cite{Ando00,Huertas-Hernando06}
\begin{equation}
 H_{\rm vib} = \frac{p^2}{2M}+ \frac{1}{2}M\omega_0^2 q^2,
\end{equation}
with CNT mass $M$, resonant frequency $\omega_0$, and the displacement $q$.

The flexural vibration of CNT QD couples to the electronic degree of freedom through
two mechanisms: 1) the influence to the tangent vector which results in spin-vibration 
coupling \cite{Rudner&Rashba}; 2) the influence to the gate capacitance which 
induces the density-vibration coupling \cite{Sapmaz03,Lassagne09,Steele09}. 
For the first case, 
the flexural vibration (along $x$-direction) changes the tangent vector $\hat{\textbf{t}}$;
and the vector becomes coordinate dependent $\hat{\textbf{t}}(z)=(\partial_{z} q/(\sqrt{1+(\partial_{z} q)^2}),
0,1/(\sqrt{1+(\partial_{z} q)^2}))$. Up to second order of the small quantity $\partial_{z} q$, one has 
\begin{equation}
\hat{\textbf{t}}(z)\backsimeq (1-(\partial_{z} q)^2/2)\hat{\textbf{z}}+\partial_{z} q \hat{\textbf{x}}.
\end{equation}
The derivative of the displacement can be obtained:
$\langle\partial_z q\rangle \sim \int_{-L/2}^{L/2}dz\,\rho(z)(df(z)/dz)$,
where $\rho(z)$ is the electron density, $f(z)$ is the dimensionless wave function form,
and $L$ is the length of the CNT. For symmetric
quantum dot, the integral vanishes for even vibration mode. This cancellation can be avoided
by introducing an asymmetric potential, considering odd vibration
mode, or confining the electron only in part of the suspended CNT\cite{Palyi12}.
We choose last choice \cite{Palyi12} for simplicity's sake and obtain 
$\langle \partial_z q \rangle\backsimeq q/L$ and $\langle (\partial_z q)^2 \rangle\backsimeq q^2/L^2$
(a constant pre-factor $\sim 1$ is dropped), the similar result can be obtained for other two choices. 
The nonlinear term in $\hat{\textbf{t}}(z)$ can be neglected since $q\ll L$.
For the second case, the gate capacitance becomes a function of the displacement $q$:
$C_g(q)=C_g^0+\partial_q C_g^0 q + \partial_q^2 C_g^0 q^2/2+\cdots$ \cite{Sapmaz03,Lassagne09,Steele09}.
Since $(\partial_q C_g^0 q)/(\partial_q^2 C_g^0 q^2 /2)\sim h/q\gg 1$ ( $h$ is the distance between
CNT and the gate), the second and higher order terms can be neglected.  
We use a capacitor model to describe the electron-electron interaction, 
and in the presence of the vibration, we have
\begin{equation}
 H_{\rm INT} = E_c (N-N_g^0)^2 - E_c\frac{2V_g}{e} \partial_q C_g^0 N q
\end{equation}
where $E_{c}=e^{2}/(2C_{\sum})$ is the Coulomb charging energy with
total capacitance $C_{\sum}=C_{L}+C_{R}+C_{g}$, $N$ is the electron number operator in QD, and
$N_g=V_g C_g^0 /2$ denotes the background charge with gate voltage $V_g$.
For $T,V_{\rm SD}\ll \Delta <E_c$ ($T$, $V_{\rm SD}$, and $\Delta$ are
temperature, source-drain voltage, and level spacing of QD), only a single energy level ($d$)
near the Fermi level is relevant. The Hamiltonian of the CNT QD can be written as
\begin{eqnarray}
H_{D}&=& E_{c}(N-N_{g}^{0})^{2}+ \frac{p^2}{2M}+ \frac{1}{2}M\omega_0^2 q^2+\frac{\Delta_{SO}}{2}\tau_{3}s_{z}\nonumber\\
 &&+\Delta_{KK'}\tau_{1}+\lambda_{D}N\, q+\lambda_{SO}\tau_{3}s_{x}\, q
\end{eqnarray}
where $N=\sum_{\sigma=\{s,\tau\}}d_{\sigma}^{\dagger}d_{\sigma}$,
$\lambda_{SO}=\Delta_{SO}/L$, and $\lambda_{D}=-E_{c}2V_{g}\partial_{q}C_{g}^{0}/e$.
The operator $d_{\sigma}$ annihilates a spin $\sigma$ electron  in the CNT QD.

The CNT QD is coupled to two CNT leads, and the Hamiltonian
for the whole system shown in Fig. \ref{fig:setup_KondoV} is
\begin{eqnarray}
H & = & \sum_{\alpha=L/R}\sum_{k}\sum_{\sigma=\{s,\tau\}}\epsilon_{k,\sigma}c_{\alpha k,\sigma}^{\dagger}c_{\alpha k,\sigma}+
         H_{D}\nonumber \\
 &  & \quad+\sum_{\alpha}\sum_{k}\sum_{\sigma=\{s,\tau\}}V_{\alpha k}\left(c_{\alpha k\sigma}^{\dagger}d_{\sigma}+h.c.\right)
\label{eq:H_KondoV}
\end{eqnarray}
where $c_{\alpha k,\sigma}$ annihilates an electron with momentum $k$ in the $\alpha$ lead with spin
$\sigma$. $V_{\alpha k}$ describes the tunneling strength between CNT QD and the $\alpha$
lead, and can be controlled by two side gate shown in Fig. \ref{fig:setup_KondoV}.

\section{Perturbation treatment for weak electron-vibration coupling}

We want to study how the electron dynamics
affect the physics of the resonator. For weak electron-vibration
coupling ($\lambda_{SO},\lambda_{D}\ll\Gamma_{e}$), we treat $\lambda_{SO}$ and $\lambda_{D}$
as small parameter. Here, $\Gamma_{e}$ indicates the electron energy
scale in the system, i.e. Kondo temperature in the Kondo regime, hybridization 
$\Gamma=\pi\rho_0 |V_L|^2+\pi\rho_0 |V_R|^2$ ($\rho_0$ is the electron density of state in the leads)
in the mixed valence regime (i.e. the energy level of the dot $\epsilon_d$ is closed to the fermi level
$|\epsilon_d|\ll \Gamma$). 

The electron vibration coupling can be written in a general form $H_{\rm e-v}=h^{(1)} q$,
where the linear coupling $h^{(1)}=\lambda_{D}N+\lambda_{SO}\tau_{3}s_{x}$
in our problem. For small coupling, the linear response theory results in
\begin{equation}
h^{(1)}(t) =  h_{0}^{(1)}(t)-\int_{-\infty}^{\infty}\alpha_{h}^{(1)}(t')q(t-t')dt'
\label{eq:linear_response}
\end{equation}
where $\alpha_{h}^{(1)}(t-t')=i\theta(t-t')\langle[h^{(1)}(t),\; h^{(1)}(t')]\rangle_{0}$ and
$\langle\cdots \rangle_0$ indicates the average for system without electron-vibration coupling.
By solving the Heisenberg equation of motion,
we can obtain the equation describing the dynamics of the resonator 
in the linear response limit \cite{dykman84}
\begin{equation}
 \ddot{q}+2\gamma \dot{q} +\omega_0^2 q =\frac{F}{M}\cos(\omega_F t)-\frac{h^{(1)}(t)}{M}
\label{eq:resonator_DEQ_q}
\end{equation}
where we include a periodic driving force $F$ with frequency $\omega_F$, and a bare damping 
$\gamma=\omega_0/Q_0$ with quality factor $Q_0$. In the limit $\gamma,\,|\omega_{F}-\omega_{0}|\,\ll\omega_{0}$,
one can analyze the problem in a rotating frame using the following
transformation \cite{dykman84}
\begin{eqnarray}
q(t) & = & u(t)\, e^{i\omega_{F}t}+u^{*}(t)\, e^{-i\omega_{F}t}\nonumber \\
\dot{q}(t) & = & i\omega_{F}\left(u(t)\, e^{i\omega_{F}t}-u^{*}(t)\, e^{-i\omega_{F}t}\right)
\label{eq:RFT}
\end{eqnarray}
Combining Eq.~(\ref{eq:linear_response}),~(\ref{eq:resonator_DEQ_q}) and ~(\ref{eq:RFT}),
and then  neglect the fast oscillating
terms (like $e^{\pm i\omega_{F}t}$, $e^{\pm2i\omega_{F}t}$, $\cdots$),
the equation of motion for the resonator becomes \cite{dykman84}
\begin{eqnarray}
&&\dot{u}=-i\Big[\omega_{F}-\omega_{0}+\frac{{\rm Re}\Big(\alpha_{h}^{(1)}(\omega_{F})\Big)}{2M\omega_{F}}\Big]u \\
&&   -\Big[\gamma+\frac{{\rm Im}\Big(\alpha_{h}^{(1)}(\omega_{F})\Big)}{2M\omega_{F}}\Big]u 
  -i\frac{F}{4M\omega_{F}}+i\frac{h_{0}^{(1)}(t)}{2M\omega_{F}}e^{-i\omega_{F}t}.\nonumber
\label{eq:resonator_DEQ_u}
\end{eqnarray}
One then obtain the damping ($\omega_F\approx\omega_0$) due to electron vibration coupling 
\begin{eqnarray}
\gamma^{e-v}&&=\gamma^{s-v}+\gamma^{d-v}= \frac{{\rm Im}\left[\alpha_{h}^{(1)}(\omega_{0})\right]}{2M\omega_{0}}\\
&&=\frac{\lambda_{SO}^{2} {\rm Im}[\chi_{\tau_{3}s_{x}}(\omega_{0})]}{2M\omega_{0}}
      +\frac{\lambda_{D}^{2} {\rm Im}\left[\chi_{N}(\omega_{0})\right]}{2M\omega_{0}},\nonumber
\label{eq:damping}
\end{eqnarray}
where $\chi_{\tau_{3}s_{x}}(\omega)$ and $\chi_{N}(\omega)$ represent 
the dynamical spin susceptibility and density susceptibility respectively,
which are the Fourier transforms of the functions 
$\chi_{\tau_{3}s_{x}}(t)=i\theta(t)\langle[\tau_{3}s_{x}(t),\tau_{3}s_{x}(0)]\rangle_{0}$
and $\chi_{N}(t)=i\langle[N(t),\; N(0)]\rangle_{0}$. 
We choose $\hbar=k_B =1$ throughout the paper. The frequency shift due to electron resonator
interaction is 
\begin{equation}
\Delta\omega^{s-v}=\frac{{\rm Re}\left[\alpha_{h}^{(1)}(\omega_{0})\right]}{2M\omega_{0}},
\end{equation}
corresponding to the real part of the sum of the dynamical spin susceptibility and density susceptibility.
The last term in Eq.~(\ref{eq:resonator_DEQ_u}) describes
the noise. 

In equilibrium, those susceptibilities are directly related to the spin 
and change fluctuations via fluctuation dissipation theorem.
The fluctuations show different behaviors in different regimes.
In the mixed valence regime ($|\epsilon_d|\ll\Gamma$), large charge and spin fluctuations
\cite{Brunner&Langreth97,Lassagne09,Steele09} (i.e. electrons hop onto and off the CNT QD)
induce large damping and frequency shift of the CNT resonator.
The damping and frequency shift effects become weaker as temperature 
decreases. If energy level $\epsilon_d$ lays in the middle of conductance valley, 
electron tunnelings are blockaded.
In low $T$ limit, this middle valley regime can be classified into
two cases: 1) The total spin of QD is a singlet, 2) the total spin is a non-singlet.
For the first case, no spin exchange coupling can be generated in low energy; and 
thus the fluctuations will be suppressed in low $T$ (with very small leftover due to
quantum mechanical co-tunneling processes). For the second case, spin exchange coupling
is generated in low $T$ and results in Kondo effects \cite{HewsonBook} when $T<T_K$
($T_K$ is called Kondo temperature). Kondo effects induce large spin fluctuations;
and therefore, we expect large damping and frequency shift effects of the resonator
in the regime $T\sim T_K$, and the effects become stronger as decreasing $T$.
Although the charge fluctuations show an enhancement due to Kondo resonance,
their value are much smaller than that of spin fluctuations \cite{Brunner&Langreth97}.
For $\lambda_{SO}\sim \lambda_D$, we can neglect the charge fluctuation 
part (i.e. the second term in Eq.~(\ref{eq:damping})).
We now focus on the spin fluctuation part in the Kondo regime in the rest of the paper.

\section{Kondo regime}
\begin{figure}[t]
\centering
\includegraphics[width=2.8in,clip]{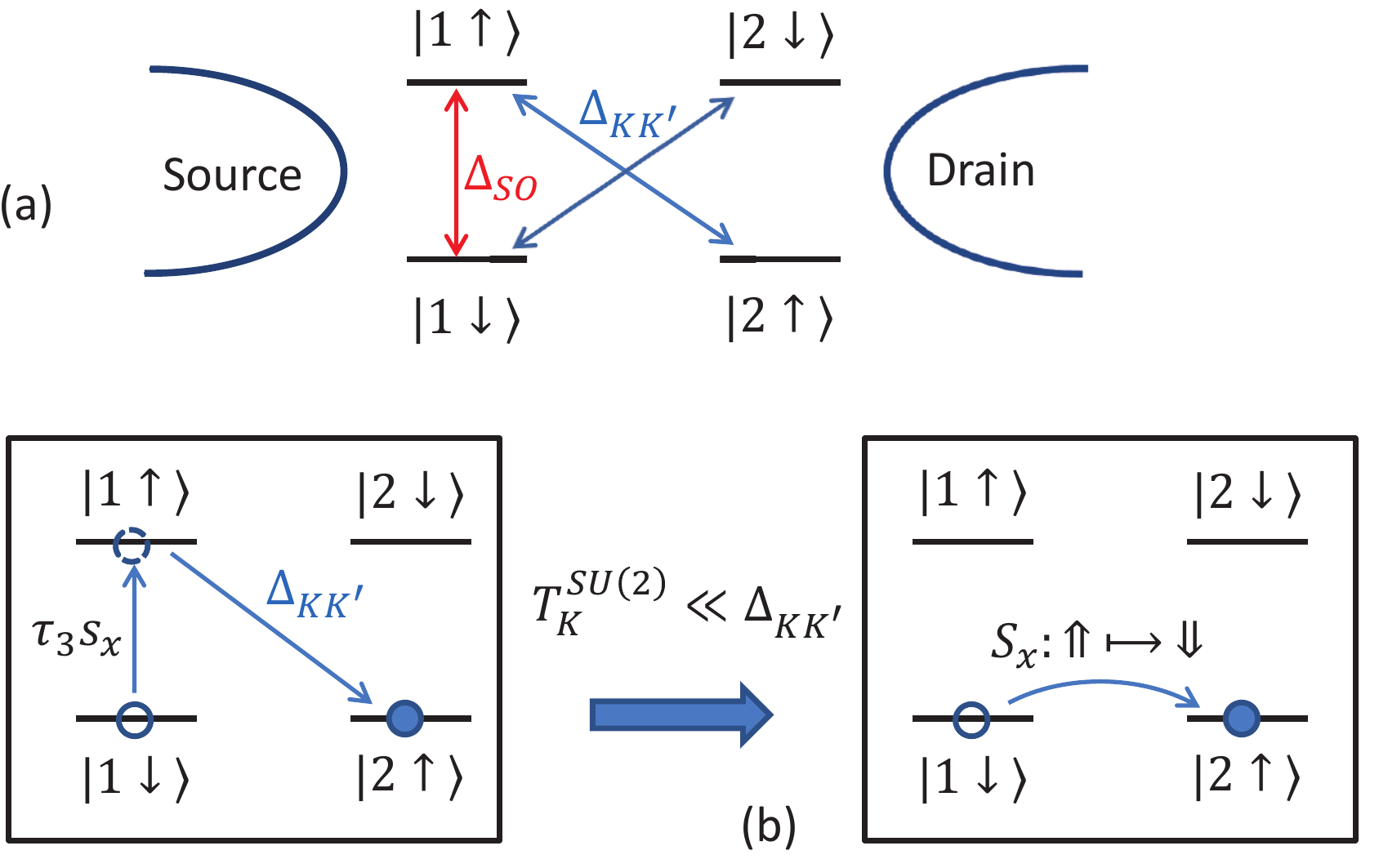}
\caption{(color online) (a) Energy splitting due to spin orbit coupling.
The blue cross arrows indicate the intervalley scatterings.
(b) In the limit $T_K^{SU(2)}\ll\Delta_{KK'}$, the operator $\tau_3 s_z$
is equivalent to the operator $S_x$ in the lower subspace in the low energy. 
The coupling becomes $\frac{\lambda_{SO}q\,\Delta_{KK'}}{\Delta_{SO}}=\lambda_{KK'}q$.
}
\label{fig:SU2Kondo}
\end{figure}

We want to calculate the the dynamical spin susceptibility $\chi_{\tau_{3}s_{x}}(\omega)$
of the system shown in Eq.~(\ref{eq:H_KondoV}) for $\lambda_{SO}=0$
and $\lambda_{D}=0$. For $\Delta_{SO}=0$ and $\Delta_{KK'}=0$, this model shows $SU(4)$ Kondo
effects \cite{Choi05,Jarillo-Herrero05} if the energy level is neither empty nor fully occupied 
(filled by 4 electrons). The large S-O interaction splits the four-fold degenerate
spin states: $(1\uparrow,1\downarrow,2\uparrow,2\downarrow)$.
Two lower energy states $(1\downarrow,2\uparrow)$ form a two-fold
degenerate subspace, and two other states form another
two-fold subspace as shown in Fig. \ref{fig:SU2Kondo} (a). 
For single occupied case, the system shows $SU(2)$ Kondo 
physics with two isospin states $(1\downarrow,2\uparrow)$ \cite{GalpinPRB10}.
For doubly occupied case, there is no Kondo effect. We have similar effects for triple
occupied and fully occupied cases. If all the parameters ($E_c$, $\epsilon_d$, and $\Gamma$) 
are the same, we have $T_{K}^{SU(2)}\ll T_{K}^{SU(4)}$ \cite{Choi05}.
The intervalley scatterings only generate the transitions $1\downarrow\leftrightarrow2\downarrow$
or $2\uparrow\leftrightarrow1\uparrow$, and thus do not affect the $SU(2)$ Kondo physics 
if $\Delta_{KK'}\ll\Delta_{SO}$ (typical experimental value: $\Delta_{SO}=370\mu eV$
and $\Delta_{KK'}=32.5\mu eV$ \cite{Kuemmeth08}).

We are interested in an experimentally realizable regime: 
$\lambda_{SO}q_{\rm amp}\ll\omega_{0}\sim T_{K}^{SU(2)}\ll T_{K}^{SU(4)}<\Delta_{KK'}<\Delta_{SO}$,
where $q_{\rm amp}$ is the amplitude of the vibration.
If the system is in the middle of the single occupied valley, two
isospin states $(1\downarrow,2\uparrow)=(\Downarrow,\Uparrow)$
along with the leads form a $SU(2)$ Kondo state. 
The Kondo effect enhances the spin-flip $\Downarrow\leftrightarrow\Uparrow$ scattering processes between two isospin
states, and their time scale (between two adjacent spin-flip events) corresponds to
$\tau_{{\rm SF}}\sim 1/T_{K}^{SU(2)}$. In the spin susceptibility, when
the operator $\tau_{3}s_{x}$ is applied on the impurity
state, the impurity state will immediately go
to the higher energy subspace: from
$1\downarrow$ to $1\uparrow$ (or from $2\downarrow$ to $2\uparrow$).
The system will finally relax to the lower energy subspace through two
possible scattering mechanism: 1) spin (real spin or isospin) exchange processes
via dot-leads hopping, 2) intervalley scattering. 
The leading relaxation processes are the intervalley scatterings 
(time scale comparison: $1/\Delta_{KK'}<1/T_{K}^{SU(4)}$).
In addition, this relaxation time is much smaller than spin-flip
time scale, i.e. $1/\Delta_{KK'}\ll 1/T_{K}^{SU(2)}$.
So, if we only consider the low energy physics $\omega\sim T_{K}^{SU(2)}$,
the operator $\tau_{3}s_{x}$ along with the fast relaxation process
just behave as the $S_{x}$ operator in the isospin
subspace $(\Downarrow,\Uparrow)$ as demonstrated in Fig. (\ref{fig:SU2Kondo}) (b),
and the vibration-spin coupling becomes $\frac{\lambda_{SO}\Delta_{KK'}}{\Delta_{SO}}S_{x}q=\lambda_{KK'}S_{x}q$ with
$\lambda_{KK'}=\Delta_{KK'}/L$.
In the low energy, the system will
exhibit the $SU(2)$ rotation symmetry. 
When considering the low energy response $\omega\sim T_{K}^{SU(2)}$,
we can safely neglect the $SU(2)$ symmetry breaking terms, and the
spin $x$ response function are exactly the same as the spin $z$
response function. Therefore, our task is reduced to the standard
problem: Calculating the response functions
\begin{equation}
\gamma^{s-v}=\frac{\lambda_{KK'}^{2}{\rm Im}\left[\chi_{S_{z}}(\omega_{0})\right]}{2M\omega_{0}},\;\,
\Delta\omega^{s-v}=\frac{\lambda_{KK'}^{2}{\rm Re}\left[\chi_{S_{z}}(\omega_{0})\right]}{2M\omega_{0}}
\label{eq:KondoDamping}
\end{equation}
for a two-fold degenerate Anderson model
\begin{eqnarray}
H & = & \sum_{\alpha,k}\sum_{\sigma}\epsilon_{k,\sigma}c_{\alpha k,\sigma}^{\dagger}c_{\alpha k,\sigma}
       +\sum_{\sigma}\epsilon_{d}d_{\sigma}^{\dagger}d_{\sigma}
        +U d_{\Uparrow}^{\dagger}d_{\Uparrow} d_{\Downarrow}^{\dagger}d_{\Downarrow}  \nonumber \\
&  & +\sum_{\alpha k \sigma}V_{\alpha k}\left(c_{\alpha k\sigma}^{\dagger}d_{\sigma}+h.c.\right)\;.
\label{eq:SU2_H}
\end{eqnarray}
where $\sigma=\{\Downarrow,\Uparrow\}$, $U=2E_c$, and 
$S_z=(d_{\Uparrow}^{\dagger}d_{\Uparrow}-d_{\Downarrow}^{\dagger}d_{\Downarrow})/2$.

\section{Damping and frequency shift due to Kondo correlation }

We will calculate the damping and frequency shift
of the CNT resonator due to electron-vibration coupling induced 
in the Kondo regime, i.e. Eq.~(\ref{eq:KondoDamping}) for
Hamiltonian in Eq.~(\ref{eq:SU2_H}). 
Although the exact result might be obtained from numerical techniques,
e.g. numerical renormalization group, real-time quantum Monte Carlo techniques, 
analytic treatments are extremely non-trivial. Since spin-flip process can occur through
the virtual processes involving either the empty state or the doubly occupied state, essential physics of
the Kondo effect will not be significantly affected if one remove
the doubly occupied state from the Hilbert space. To study the Kondo
physics, one then can use the standard simplification, i.e. study
an infinite-U (interaction) Anderson model. This model can be solved by using a standard 
non-crossing approximation (NCA) \cite{Coleman84,BickersPRB87,HewsonBook},
which is exact for large degeneracy limit. But it can capture
the essential physics for energy above a pathology scale $\sim10^{-1}-10^{-2}T_{K}$
\cite{BickersRMP87,HewsonBook} even for our $2$-fold degenerate case.

\begin{figure}[t]
\centering
\includegraphics[width=3.2in,clip]{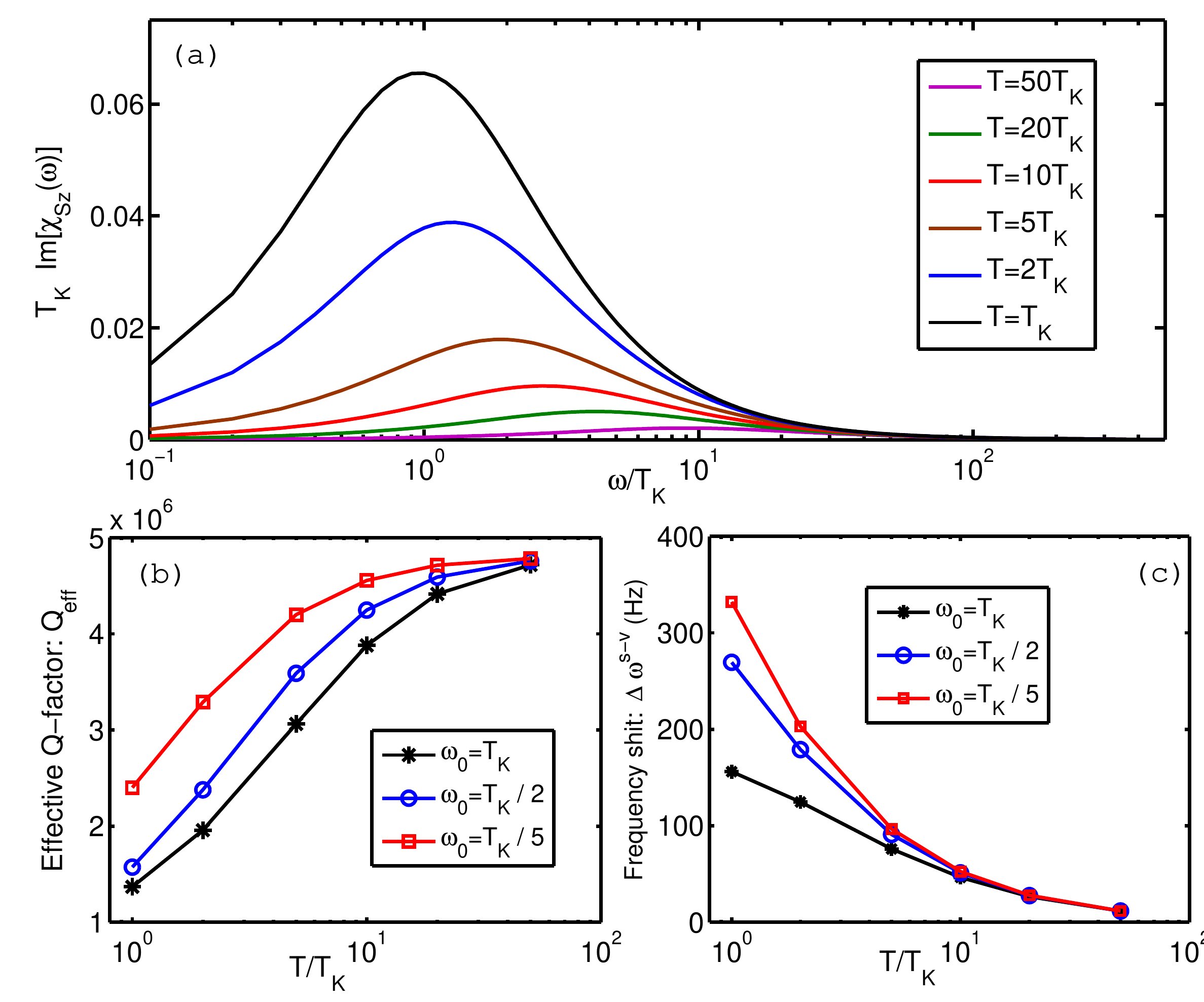}
\caption{(color online) (a) The imaginary part of the dynamical spin susceptibility 
as a function of the frequency $\omega$ for $D=5.0$, $\Gamma=\Gamma_L+\Gamma_R=0.5$,
and $\epsilon_d=-1.5$. From bottom to top: $T=50T_K,\,\cdots,\,T_K$.
(b) The effective quality factor, (c) the frequency shift
of the CNT resonator. The parameters are shown in the text.
}
\label{fig:MagSUS_QF_FS}
\end{figure}

The Green function of problem can be obtained by solving a set of coupled integral equations iteratively
\cite{Coleman84,BickersPRB87,HewsonBook} (also refer to Apendix \cite{app} for more details about the calculation of dynamical spin susceptibility in Kondo regime). Fig. \ref{fig:MagSUS_QF_FS} (a) shows
the imaginary part of the dynamical spin susceptibility $T_K \rm{Im}[\chi_{S_z}(\omega)]$ as
a function of the frequency $\omega/T_K$ for different temperatures.
We choose the parameters: electron lead band width $D=5.0$, $\Gamma=\Gamma_L+\Gamma_R=0.5$,
and $\epsilon_d=-1.5$ such that the Kondo temperature $T_K\approx 0.01$ 
(this is obtained from the width of the impurity spectrum function).
As expected \cite{Coleman84,BickersPRB87,Brunner&Langreth97,HewsonBook}, a peak maximum appears in 
the susceptibility spectrum; and it is shifted to lower frequency and approaches 
$\omega=T_K$ as $T$ is lowered. The real part of the susceptibility spectrum is related
to its imaginary part through Kramers-Kronig relation.
We then calculate the effective quality factor $Q_{\rm eff}$, which is given as
\begin{equation}
 \frac{\omega_0}{Q_{\rm eff}} = \frac{\omega_0}{Q_{0}} + \gamma_{s-v},
\end{equation}
and the frequency shift $\Delta\omega_{s-v}$.
We choose the following experimental realizable parameters 
\cite{Poot&vanderZantReview,Moser14,Jespersen11}: bare quality factor $Q_0=4.8\times 10^6$,
suspended CNT length and mass $L=1.8 \mu m$ and $M=4.4\times 10^{-21}kg$, resonant frequency
$\omega_0=2\pi f_0=2\pi \times 55.6 MHz$, $\Delta_{KK'}=200\mu eV$ (note that the intervalley scattering comes from 
disorder and their value fluctuates from device to device, Jespersen, et al. reported the value as large as $450\mu eV$.), and
Kondo temperature $T_K=T_K^{SU(2)}=\omega_0$, $2\omega_0$, and $5\omega_0$. The effective electron-vibration coupling
is estimated in their zero-point motion state
$\lambda_{SO}q_0 = \Delta_{SO}/(L \sqrt{2M\omega_0})\backsimeq 0.0028\omega_0\ll \omega_0$,
which is small enough to justify the perturbation method. 
The effective quality factor and the frequency shift of the CNT resonator are shown 
in Fig. \ref{fig:MagSUS_QF_FS} (b) and (c). It is clear that as the $T$ is lowered
and approaches the Kondo regime $T\sim T_K$, the strong spin-flip scatterings  
between CNT QD resonator and their leads, along with
the spin-orbit interaction, induce a large damping effect and thus decrease the 
effective quality factor dramatically.
The effective resonant frequency of the resonator is also affected by Kondo effect:
The frequency shift increases as $T$ decreases. 

The authors grateful to Y.Liu, A. Levchenko, M.I. Dykman and H.U.Baranger for valuable discussions.
The authors acknowledge support from Thousand-Young-Talent program of China, and the startup grant from State Key Laboratory of Low-Dimensional Quantum Physics and Tsinghua University.

\appendix

\section{Dynamical spin susceptibility}\label{app}

In this appendix, we will calculate the Kondo induced damping and frequency shift, i.e.
Eq. (13) for the Hamiltonian shown in Eq. (14).
Since spin-flip process can occur through
both the virtual process involving the empty state and the virtual
process involving the doubly occupied state, essential physics of
the Kondo effect will not be significantly affected if one remove
the doubly occupied state from the Hilbert space. To study the Kondo
physics, one then can use the standard simplification, i.e. study
an infinite-U (interaction) Anderson model instead of the finite-U
Anderson model shown in Eq. (14). The Hamiltonian for
N-fold degenerate model can be written as
\begin{eqnarray}
H&=&\sum_{m=1}^{N}\sum_{\alpha k}\epsilon_{\alpha k}c_{\alpha km}^{\dagger}c_{\alpha km}+
\epsilon_{d}\sum_{m=1}^{N}\hat{N}_{m}\nonumber\\
&& +\sum_{\alpha k}\sum_{m}V_{\alpha k}\left(c_{\alpha km}^{\dagger}F_{m}+h.c.\right),
\end{eqnarray}
where $\hat{N}_{0}=|0\rangle\langle0|$, $\hat{N}_{m}=|m\rangle\langle m|$,
and $F_{m}=|0\rangle\langle m|$ ($\langle0|$ represents the empty
state and $\langle m|$ represents the occupied state). This model
can be solved using a standard non-crossing approximation 
(NCA) \cite{Coleman84,BickersPRB87,HewsonBook} which neglect all the crossing diagram.
This method is justified for large N limit, but still can capture
the essential physics for energy above a pathology scale $\sim10^{-1}-10^{-2}T_{K}$
\cite{BickersRMP87,HewsonBook} even for $N=2$. The NCA
method gives coupled integral equations \cite{Coleman84,BickersPRB87,HewsonBook} 
for the empty state propagator $G_{0}$ and occupied
state propagator $G_{m}$ 
\begin{eqnarray}
G_{0}(\omega) & = & \frac{1}{\omega-\Sigma_{0}(\omega)}\\
G_{m}(\omega) & = & \frac{1}{\omega-\epsilon_{d}-\Sigma_{m}(\omega)}\\
\Sigma_{0}(\omega) & = & \frac{N\Gamma}{\pi}\int_{-D}^{D}\frac{f(\omega)}{\omega+\epsilon-\epsilon_{d}-\Sigma_{m}(\omega+\epsilon)}d\epsilon\\
\Sigma_{m}(\omega) & = & \frac{\Gamma}{\pi}\int_{-D}^{D}\frac{1-f(\omega)}{\omega-\epsilon-\Sigma_{m}(\omega-\epsilon)}d\epsilon
\end{eqnarray}
Here $\Gamma=\Gamma_{L}+\Gamma_{R}$ ($\Gamma_{\alpha}=\pi|V_{\alpha}|^{2}\rho$)
and $D$ is the half band width of the system. The coupled integral
equation can be simply solved using iteration method. The empty-state
spectrum and the occupied state spectrum are
\begin{eqnarray}
\rho_{0}(\omega) & = & -\frac{1}{\pi}{\rm Im}G_{0}^{R}(\omega)\\
\rho_{m}(\omega) & = & -\frac{1}{\pi}{\rm Im}G_{m}^{R}(\omega)
\end{eqnarray}
Then, the physical observables, e.g. full impurity spectral function
and the dynamical spin susceptibility can be obtained
\begin{eqnarray}
A_{d}(\omega) & = & \frac{1}{Z_{f}}(1+e^{-\beta\omega})\int_{-\infty}^{\infty}d\epsilon\; e^{-\beta\epsilon}\rho_{0}(\epsilon)\rho_{m}(\epsilon+\omega)\nonumber\\
{\rm Im}\left[\chi_{S_{z}}(\omega)\right] & = & \frac{N\, j(j+1)}{3}\frac{\pi}{Z_{f}}\int_{-\infty}^{\infty}d\epsilon\; e^{-\beta\epsilon}\rho_{m}(\epsilon)\nonumber\\
 & & \times\left[\rho_{m}(\epsilon+\omega)-\rho_{m}(\epsilon-\omega)\right]
\end{eqnarray}
where $j=(N-1)/2$ and 
\begin{equation}
Z_{f}=\int_{-\infty}^{\infty}d\epsilon\; e^{-\beta\epsilon}\left[\rho_{0}(\epsilon)+\sum_{m}\rho_{m}(\epsilon)\right].
\end{equation}

\begin{figure}[t]
\centering
\includegraphics[width=3.2in,clip]{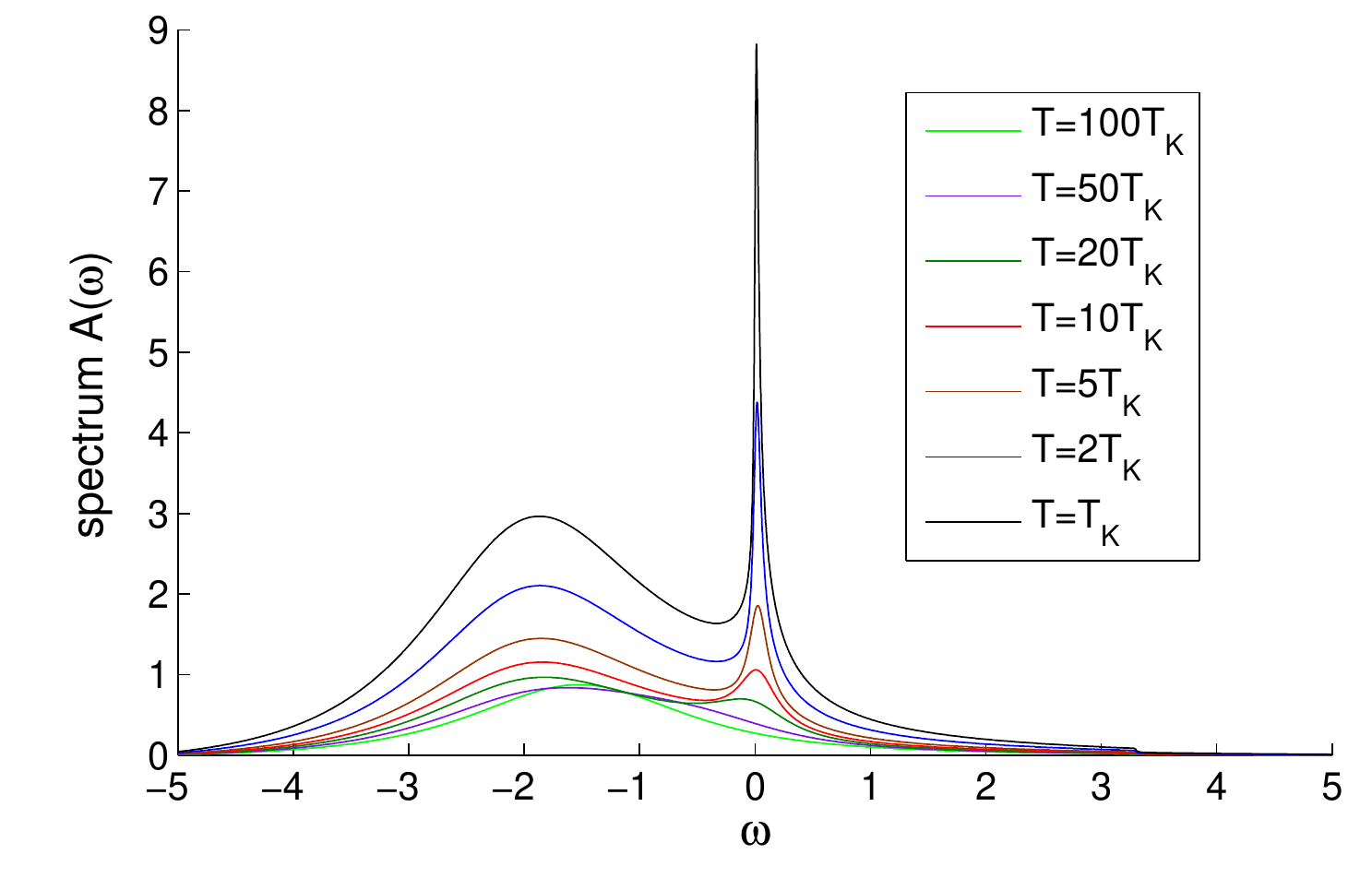}
\caption{The impurity spectral function for different temperature.
}
\label{fig:spectrum_N2}
\end{figure}

For our problem, the degeneracy is $N=2$. We did a calculation for
$D=5.0$, $\Gamma=0.5$, and $\epsilon_{d}=-1.5$, so that the Kondo
temperature is $T_{K}\sim0.01$ (this can be directly obtained from
the width of the Kondo resonance of the impurity spectrum $A_{d}(\omega)$).
We test the impurity spectral function for different temperature shown
in Fig. \ref{fig:spectrum_N2}, we can see that the narrow
Kondo resonance around $\omega=0$ is developed for low temperature.
We then obtain the dynamical spin susceptibility numerically as shown in Fig.3 of the main text.

\bibliography{KondoV}

\end{document}